\documentclass[aps,prl, twocolumn,superscriptaddress]{revtex4}
\usepackage{graphicx}

\def\psit{{\tilde \psi}}
\def\Jt{{\tilde J}}
\def\zetat{{\tilde \zeta}}

\begin{document}\title{Perturbative approach to the nonlinear saturation of the tearing mode for any current gradient }

\author{ N. Arcis }
\affiliation{Association EURATOM-CEA, CEA/DSM/DRFC, CEA Cadarache,
13108 St Paul-lez-Durance, France}
\email  {E-mail : Nicolas.Arcis@cea.fr.}

\author{D.F.~Escande}
\affiliation{ UMR 6633 CNRS--Universit{\'e} de Provence,
case 321, Centre de Saint-J{\'e}r{\^o}me,
F-13397 Marseille cedex 20}
\email  {E-mail : escande@up.univ-mrs.fr.}

\author{M.~Ottaviani}
\affiliation{Association EURATOM-CEA, CEA/DSM/DRFC, CEA Cadarache,
13108 St Paul-lez-Durance, France}
\email {Maurizio.Ottaviani@cea.fr}

\date{\today}

\begin{abstract}Within the traditional frame of reduced MHD, a new rigorous perturbation expansion provides the equation ruling the nonlinear growth and saturation of the tearing mode for any current gradient. The small parameter is the magnetic island width $w$. For the first time, the final equation displays at once terms of order $w \ln(1/ w)$ and $w$ which have the same magnitude for practical purposes; two new $O(w)$ terms involve the current gradient. The technique is applicable to the case of an external forcing. The solution for a static forcing is computed explicitly and it exhibits three physical regimes.

\end{abstract}


\maketitle

Many out of equilibrium macroscopic media display bifurcations leading to the build up of macroscopic structures. In magnetized plasmas, magnetic islands are such structures. They can develop in the presence of a current inhomogeneity through the instability of the tearing mode \cite{Furth} which produces magnetic reconnection. This mode corresponds to a global magnetic perturbation  that is resonant in a spatial region where its wave-number is perpendicular to the magnetic field. The magnetic island occurs in this region and stretches along the wave-number direction. The nonlinear saturation of a tearing magnetic island is the simplest instance of magnetic self-organization in a plasma. Tokamak operation avoids the formation of such islands, since they degrade confinement. In the reversed-field pinch (RFP), the occurrence of several magnetic islands leads to magnetic chaos spoiling confinement, but the formation of a single magnetic island is desirable, since it should provide good magnetic flux surfaces and a laminar dynamo. Therefore, a correct description and understanding of the nonlinear tearing mode is both important for thermonuclear fusion and for advancing the theory of plasma self-organization.

The nonlinear tearing mode is classically described by applying  resistive reduced  magnetohydrodynamics (RRMHD) to the model of a static plasma slab, in the limit of small dissipation \cite{Ruth,White,Thya,MP,EO}. The magnetic island region is considered as a boundary layer whose nonlinear features are dealt with,  while the outer region is adequately described by linear theory only. The inner and outer solutions are then matched asymptotically. Rutherford \cite{Ruth} showed that the Navier-Stokes equation of RRMHD reduces in the nonlinear regime to a mere Grad-Shafranov equation, and proved a linear growth of the island width $w$ to follow the exponential growth of the linear regime. The island saturation was dealt with approximately in Refs. \cite{White}, and rigorously in Ref. \cite{Thya}, for a large enough current gradient in the island region, and in Refs. \cite{MP,EO} for a vanishing gradient.

Within the above classical setting, this Letter brings important novelties both in the method and in the final results. A new rigorous perturbation expansion using the magnetic island width $w$ as a small parameter is applied in the inner region where no assumption is made on the current gradient. This technique is in principle workable at any order, and is applicable to the case of an external forcing in a plasma with a velocity profile. The explicit solution is provided and discussed for the case of static forcing. Both uniform and non uniform electric field profiles are considered. The final equation describing the nonlinear island growth and saturation displays together the known term of order $w \ln(1/w)$ \cite{Thya} as well as a known \cite{MP,EO} and two new terms of order $w$. Both orders have the same importance for physical applications, and the first neglected term involves a factor $w^2$, which makes it physically smaller. Furthermore, the asymptotic matching reveals a jump of magnetic flux in the inner domain.

We use the  2D RRMHD equations in the
$(x,y)$ plane that is perpendicular to the magnetic field on the resonant surface
\begin{eqnarray}
  \partial_t \nabla^2 \varphi + [\varphi,\nabla^2 \varphi]
  &=&
  \nu \nabla ^4 \varphi-[\psi,J]
   \label{MHD1}
  \\
  \partial_t \psi + [\varphi,\psi]
  &=&
  \eta (J_{\rm eq} - J)
  \label{MHD2}
\end{eqnarray}
 where $\psi$ is the magnetic flux function, 
$\varphi$ is the electric potential  and plays the role of
the stream function, $\eta$
and  $\nu$ are the resistivity and viscosity which  may be weakly $x$-dependent, $J = - \nabla^2 \psi$ is the current density and $J_{\rm eq}$
its equilibrium value. $[A,B]\equiv
\partial_xA\,\partial_yB-\partial_yA\,\partial_xB$ is the 2D
Jacobian or Poisson bracket. Units are chosen such that $J_{\rm eq}(0)=1$. Furthermore, 
following Refs. \cite{Thya} and \cite{EO}, it is useful to rescale the electric potential as $\varphi \rightarrow \eta_0 \varphi$, where $\eta_0=\eta(0)$.
This makes explicit the fact that the vorticity and the viscosity terms in Eq.~\ref{MHD1} are proportional to $\eta^2$ and $\eta \nu$ respectively. These two terms can then be dropped by 
assuming the island width to be larger than the visco-resistive and resistive lengths \cite{Ruth}. Equations  ~\ref{MHD1}-\ref{MHD2} thus  become
\begin{eqnarray}
\left[\psi,J\right] & = & 0
\label{psiJ}
\\
\left[\varphi,\psi\right] & = & \eta (J_{\rm eq}-J) 
\label{phipsi}
\\
J & = & -\Delta \psi \  
\label{J}
\end{eqnarray}
In the following, two different models have been considered: model A,
in which the resistivity is constant, $\eta=1$, and $ J_{\rm eq}$ is given
as a power series in $x$, $ J_{\rm eq}=1+\sum_{i=1}^{\infty}b_i x^i$, and 
model B, in which $\eta$ is not constant, $\eta=1+\sum_{i=1}^{\infty} a_i x^i$,
but  the product $\eta J_{\rm eq}$ is constant (uniform electric field). In this instance, the coefficients $b_i$'s
can be expressed in terms of the $a_i$'s (in particular,
$a_1=-b_1$ and $a_2=b_1^2-b_2$) In  both cases,
the origin is chosen to be the location of a null of the equilibrium field around which a magnetic island develops. We thus  take the equilibrium flux function to be $\psi_0\sim -x^2/2- b_1 x^3/6 -b_2 x^4/12$ in a neighborhood of the origin, where '$\sim$' is used throughout the Letter with the meaning "equals plus higher order terms". 

Small island solutions of the system of Eqs \ref{psiJ}-\ref{J} are conveniently obtained with the technique of asymptotic matching. We consider the nonlinear saturation of a tearing mode with wavenumber $k$. Let $\delta\equiv  |\psi_{\rm out 1}- \psi_0|^{1/2}$, where $\psi_{\rm out 1}$ is the dominant part of the outer solution incorporating harmonics 0 and 1 of its Fourier expansion in $ky$. We look for two classes of asymptotic solutions, one valid in the outer region such that $|x|>>\delta$, and one valid in the inner region (the island region) $|x|\approx\delta$. Matching in the overlapping region $\delta <<|x| <<1$,
where both expansions are valid, then allows to determine the free parameters of the problem, and in particular the expansion parameter $\delta$. 

In view of the matching with the inner solution, it is convenient to introduce the stretched coordinate $\xi\equiv x/\delta$, the angle $\chi \equiv ky$, and the scaled flux function $\zeta \equiv -\psi/\delta^2$. With these definitions, the outer solution is \cite{Thya}: 
\begin{eqnarray} 
\label{psiout}
\zeta_{out} & \sim&
\xi^2/2-\cos{\chi}
-\delta \ln{\delta} \, b_1\xi\cos{\chi}
\nonumber \\
& &
+\delta [b_1\xi^3/6-(\Sigma'\xi/2+\Delta'|\xi|/2+b_1\xi\ln{|\xi|}) \cos{\chi}]
\nonumber
\\
& &
 -  \delta^2\ln{\delta} \, (b_1^2/2)\xi^2\cos{\chi}
\nonumber
\\
& &
+ \delta^2 [b_2\xi^4/12- \{\Sigma'b_1\xi^2/4+\Delta'b_1|\xi|\xi/4
\nonumber
\\
& &
+(k^2/2+b_2-b_1^2)\xi^2+
(b_1^2/2) \xi^2 \ln{|\xi|}\}\cos{\chi}]
\end{eqnarray}
where $\Delta'$, the usual tearing mode stability index, is the jump of the logarithmic derivative of $\psi_{\rm out 1}$ at $x=0$, and $\Sigma'$ is the sum of the right and left values of this derivative.
 
We now proceed to the nonlinear description of the inner region. Since the saturated mode amplitude corresponds to a bifurcation ruled by the stability index $\Delta'$, our analysis will show that $\delta$ depends on $\Delta'$. First, we need to define a suitable
ordering of the fields. Since we are interested in 
tearing modes in the small island limit, it is appropriate to use the so-called
constant-$\psi$ approximation \cite{Furth}. Denoting by $\psit$ the perturbed flux function and by $\psit'$ its x-derivative, we assume that $\psit$ varies little in the island region, $\delta \psit'/\psit <<1$. Since the perturbed current
is at most of order $\psit'/\delta$ and $\psit$ itself is of order
$\delta^2$, one concludes that $J=1$ to leading order in the island region. Physically,
this means that a tearing mode island does not alter appreciably the equilibrium current. At zeroth order in $\delta$, Eq. \ref{J} shows that $\zeta_0$, the leading order of $\zeta$,  satisfies $\partial_{\xi}^2\zeta_0=1$, which implies $\zeta_0=\xi^2/2-\cos{\chi}$ upon matching with (\ref{psiout}). Moreover, Eq. \ref{phipsi} shows that $\varphi$ is at most of order $1$. We thus rewrite Eqs. \ref{psiJ}-\ref{J}, for both models, in terms of $\zetat\equiv\zeta-\zeta_0$, $\Jt \equiv J-1$, $\varphi$, and of the variables $(\xi,\chi)$
\begin{eqnarray}
[\zeta_0,\Jt ] & = & -[\zetat,\Jt ]
\label{Jt}
\\
k\delta [\zeta_0,\varphi ] & = &
- k\delta [\zetat,\varphi ]+\eta(J_{\rm eq}-1)-\eta\Jt
\label{phit}
\\
\partial_{\xi}^2\zetat & = & -\delta^2k^2\partial_{\chi}^2(\zeta_0+\zetat)+\Jt
\label{zetat}
\end{eqnarray}
where the Poisson brackets are now taken with respect to $(\xi,\chi)$. 
A few remarks are now in order. First, we note that
whereas Eqs.~\ref{Jt}-\ref{zetat} can be solved in powers of $\delta$, logarithmic contributions
 of the form $\delta^n (\ln{\delta})^m$ eventually appear due to the matching requirement with the outer solution (\ref{psiout}). Furthermore, inspection of the structure of the equations shows that the exponent in $\ln{\delta}$ is bounded by the exponent in $\delta$: $m \leq n$. We therefore make the following most general perturbation expansions
\begin{eqnarray}
\zetat & \equiv & \sum_{n,m,m \leq n}\delta^n(\ln{\delta})^m\zeta_{nm} 
\label{zetat_exp}
\\
\Jt & \equiv & \sum_{n,m,m \leq n}\delta^n(\ln{\delta})^m j_{nm}
\label{Jt_exp}
\\
\varphi & \equiv & \varphi_0+\sum_{n,m,m \leq n}\delta^n(\ln{\delta})^m  \varphi_{nm}
\label{phi_exp}
\end{eqnarray}
Finally, we note that the structure of the lowest order linear operator occurring in Eqs. \ref{Jt}-\ref{zetat}, $[\zeta_0,A]$,  makes it convenient to write the equations  in terms of the new pair of independent variables $(\zeta_0,\chi)$. The Poisson bracket is then changed  from $[A,B]$ into $\xi [A,B]$, where $[A,B]$ is now defined in terms of  $(\zeta_0,\chi)$ and $\xi$ is meant as a double-valued function of $(\zeta_0,\chi)$, $\xi=\pm [2(\zeta_0+\cos \chi)]^{1/2}$. All the equations from now on are understood in terms of these new variables. 
$\zeta_0$ identifies  magnetic  flux surfaces  to the lowest significant order in the perturbation expansion. In particular,  $\zeta_0=1$ corresponds to the separatrix. In the following, we will make systematic use, for any function $A$, of the identity $\int_{C_{\zeta_0}}\partial_{\chi}A\, d\chi = 0$ where $C_{\zeta_0}$ identifies a (lowest order) flux surface.
The term by term derivation of the various contributions to the series \ref{zetat_exp}-\ref{phi_exp}, up to ${\rm O}(\delta^2)$, is now outlined.
 
{\it Order} $\delta \ln{\delta}$. Using Eqs.~\ref{Jt}-\ref{zetat}
 and matching with (\ref{psiout}) immediately yields  $j_{11}=0$ and
 $\zeta_{11}=-b_1\xi \cos{\chi}$. 

{\it Order} $\delta$. To this order Eq.~\ref{Jt} is simply $\partial_{\chi}j_{10} = 0$, which gives
$j_{10}=J_1(\zeta_0)$. Writing (\ref{phit}) to the same order then
yields
\begin{equation} \label{10}
k\partial_{\chi}\varphi_0=b_1-J_1 / \xi
\end{equation}
Integrating (\ref{10}) along $C_{\zeta_0}$ yields
\begin{equation} \label{j1}
J_1=2\pi
b_1\alpha_{\zeta_0}\left/\int_{C_{\zeta_0}}\xi^{-1}d\chi\right.
\end{equation}
where $\alpha_{\zeta_0}=1$ for $\zeta_0>1$ and
$\alpha_{\zeta_0}=0$ otherwise.
Integration of Eq.~\ref{zetat} gives 
\begin{equation} \label{z1}
\zeta_1=\xi\int_1^{\,\zeta_0}\frac{J_1}{\xi}\,dx-\int_1^{\,\zeta_0}
J_1\,dx+\alpha(\chi)\xi+\beta(\chi) 
\end{equation}
where $\alpha(\chi)$ and $\beta(\chi)$ will be determined later by the matching conditions.

{\it Order} $\delta^2 (\ln{\delta})^2$. Equations are trivially satisfied with a vanishing term  to this order.

{\it Order} $\delta^2 \ln{\delta}$. Equation~\ref{Jt} gives the
equation
\begin{equation} \label{211}
-J_1'\partial_{\chi}
\zeta_{11}+\partial_{\chi}j_{21}=0
\end{equation}
whose solution  is 
$j_{21}=J_1'\zeta_{11}+J_{21}(\zeta_0)$ where
$J_{21}(\zeta_0)$ has yet to be determined. Using this result together with 
(\ref{phit})  and (\ref{10}) yields 
\begin{equation} \label{213}
J_{21}/\xi=k\partial_{\chi}(\zeta_{11}\partial_{\zeta_0}
\varphi_0-\varphi_{11})+ b_1^2\cos{\chi}/\xi.
\end{equation}
Integrating (\ref{213}) along $C_{\zeta_0}$ eventually gives
\begin{equation} \label{j21}
J_{21}=b_1^2\int_{C_{\zeta_0}}\frac{\cos{\chi}}{\xi}\,d\chi \left/
\int_{C_{\zeta_0}}\xi^{-1}d\chi\right.
\end{equation}
Finally, we solve for $\partial_{\xi}\zeta_{21}$ thanks to Eq. \ref{zetat}
\begin{equation}
\label{z21}
\partial_{\xi}\zeta_{21} =  -b_1J_1(\zeta_0)\cos{\chi}+\int_{-\cos{\chi}}
^{\,\zeta_0}\frac{J_{21}}{\xi}\,dx+\gamma(\chi) 
\end{equation}

{\it Order} $\delta^2$. Eq.~\ref{Jt} yields
\begin{equation} \label{21}
\partial_{\chi}j_2= J_1'\partial_{\chi} \zeta_1\end{equation}
whose solution is $j_2=J_1'\zeta_1+J_2(\zeta_0)$.  Proceeding exactly as with $J_{21}$, we get the following expressions for $J_2$ and $\partial_{\xi}\zeta_2$
\begin{equation} \label{j2}
J_2\!=\!\!\frac{\int_{C_{\zeta_0}}\left(J_1\partial_{\zeta_0}\!\{\frac{\zeta_1}{\xi}\}\!
+b_1\{\lambda J_1\! -\partial_{\zeta_0}\zeta_1\}+(b_2 - \lambda b_1^2)\xi\right)\!d\chi}{\int_{C_{\zeta_0}}
\xi^{-1}d\chi}
\end{equation}
\begin{equation} \label{z2}
\partial_{\xi}\zeta_2\!=\!\frac{J_1\zeta_1}{\xi}+\int_{-\cos{\chi}}^{\,\zeta_0}\!\!\left(\frac{J_2}{\xi}- J_1\partial_{x}
\{\frac{\zeta_1}{\xi}\}\right)
\!dx-k^2\xi\cos{\chi}+\theta(\chi)
\end{equation} 
where $\lambda =0$ for model A, and $\lambda =1$ for model B.

{\it Matching}. From (\ref{psiout}) and (\ref{z1}), the matching condition gives $\beta(\chi)=0$ and 
\begin{eqnarray}
\label{alphachi}
\alpha(\chi)&=&b_1-\cos{\chi}\left(\frac{\Sigma'+b_1\ln{2}}{2}\right.
\nonumber
\\
&&\!\!\!\!\!\!\!\!\left. + \frac{1}{\pi}\int_{-\pi}^{\pi}
\left[\int_1^{\infty}\{\frac{J_1}{\xi}+b_1\frac{\cos{y}}{2x}
\}dx\right]\cos{y}\,dy\right)\end{eqnarray}
This condition still leaves the $\Delta'$ term unmatched. Indeed, it is impossible to include it in Eq.~\ref{alphachi}, otherwise $\zeta_1$ would have a jump  at $\xi=0$ whereas inner solutions must be sufficiently smooth. We conclude that the $\Delta'$ term
must be matched  by higher order terms in the inner perturbation expansion. It is then convenient to  expand $\Delta'$ as  $\Delta'\sim\delta\ln{\delta}\Delta_{11}'+\delta\Delta_1'$.

We now proceed to the next order. Upon matching (\ref{z21}) with Eq. \ref{psiout},  one readily obtains $\gamma(\chi)=0$ and 
\begin{equation} \label{D11}
\Delta_{11}'=-\frac{2}{\pi}\int_{-\pi}^{\pi}\left(\int_{-\cos{\chi}}
^{\,\infty}\frac{J_{21}}{|\xi|}\,dx\right)\cos{\chi}\,d\chi\equiv -\ell b_1^2
\end{equation}
where $\ell$ can be computed numerically and is approximately $1.64$, which is the result  already obtained in  \cite{Thya}.

The next step of our calculation is the matching of (\ref{z2}). One proceeds as for the previous order, but the calculation is more lengthy. $\partial_{\xi}\zeta_2$ is split into a contribution diverging for large $\xi$'s and a finite term which after matching with the outer solution yields
\begin{eqnarray} \label{D1}
\Delta_1'\!\!&=&\!\!2\!\lim_{\zeta_0\rightarrow\infty}\!\!\left\{\frac{1}{\pi}\int_{-\pi}^{\pi}\!\!d\chi \cos{\chi}\!
\int_{-\cos{\chi}}^{\,\zeta_0}\!\!\left(
J_1\partial_{x}\left(\frac{\zeta_1}{|\xi|}\right)-
\frac{J_2}{|\xi|}\right)dx\right. \nonumber \\
& &\left. +\left(\frac{b_1^2}{6}+a_2\right)\sqrt
{2\zeta_0}\right\}
\end{eqnarray}

We now reintroduce time dependance and use (\ref{D11}) and (\ref{D1}) to  give our final result in terms of the island width $w\equiv 4\delta$:
\begin{equation}
 \label{result}
\eta_0^{-1} \partial_t w\sim \frac{2\Delta'}{\ell}+\frac{b_1^2}{2} w\ln{w}+\left( \frac{b_1\Sigma'}{4}-\mu \, b_1^2+b_2\right)w
\end{equation}
where numerical integration gives $\mu \approx  2.2$ for model A and $\mu \approx 2.4$ for model B. 
 The first term on the right hand side was already derived in \cite{Ruth},  the second one in \cite{Thya}, and the $b_2 w$ term in \cite{MP,EO}. If $b_1 = 0$, this formula predicts a saturation only for $b_2 < 0$. In the opposite case, there is no saturation with a small island. Furthermore, we notice that the $\Sigma'$ parameter enters this equation, which means that the contribution of order $\delta$ may be important even for a small $b_2$ if $\Sigma'$ is large enough.  Furthermore $b_1 \Sigma'$ may be positive, and thus destabilizing. 

Equation \ref{result} was obtained by matching the first Fourier harmonic in $\chi$. Matching the zeroth order harmonic at the same orders in $\delta$ brings in interesting physics too. At order $\delta$, the matching to Eq.~\ref{psiout} is possible but for a residual contribution of the form $\mp\Omega$, where $\Omega\equiv\int_1^{\infty}(J_1-b_1\sqrt{2x})\,dx-2b_1\sqrt{2}/3$. This is no problem, since one still has the freedom to add constant contributions to the flux function (\ref{psiout}) that can be different in each side of the outer region. Physically, this means that the development of a magnetic island produces a change of total magnetic flux of magnitude $2 \delta^3 \Omega$. At order $\delta^2$, the zeroth Fourier harmonic of $\partial_{\xi}\zeta_2$ is $ \mp\lambda b_1\Omega$. In
order for this term to be matched, we must
include a new contribution to (\ref{psiout}) of the form $\mp \lambda
\delta^4b_1\Omega\xi$, which is allowed since
it complies with Eq. (\ref{psiJ})-(\ref{J}). Note that this contribution vanishes for model A, as a consequence of the additional
physical constraint of total current conservation. In cylindrical geometry, the jump of magnetic flux corresponds to a flux production in the central part of the plasma by the nonlinear tearing mode. This effect was observed in a RFP \cite{Verhage}, and contributes to the toroidal field reversal, because of the global toroidal flux conservation \cite{PPCF00}.

In the presence of a static forcing, one has to add a solution of the inhomogeneous problem, which is conveniently chosen in a unique way by setting it to zero at $x=0$ \cite{HK}. Assuming that forcing is applied at the large-$x$ boundary, this solution has the form $\psi_{1f}(x,y)=0$ for $x<0$ and  $\psi_{1f}(x,y)=A_f f(x)\cos{ky}$ for $x>0$,  where $f(x)$ is the solution of the linear ideal MHD equation that satisfies $f(0)=0$, $f'(0)=1$, and $A_f$ is a parameter uniquely determined by the forced boundary conditions which can be taken positive in full generality. In the matching region, the complete solution is the sum of the unforced solution (\ref{psiout}) and of  the forced solution $\psi_{f}$,  where 
\begin{equation}
\label{psiforc}
\psi_{f} \sim  -\frac{A_f}{\delta^2}\left(\delta\frac{\xi+|\xi|}{2}+\delta^2 b_1 \frac{\xi^2+|\xi|\xi}{4}\right)\cos{\chi}
\end{equation}
Moreover, a phase difference $\phi$ between the forced and the unforced solutions can be taken into account by substituting $\cos{\chi}$ with $\cos{(\chi+\phi)}$ in (\ref{psiout}). Matching is then done separately on the $\cos{\chi}$ and  $\sin{\chi}$ components. The outcome is a set of coupled evolution equations for the island width and for the phase. One finds that fixed points occur for $\phi=0$ and $\phi=\pi$. For $\phi=0$, there is only one fixed point which is stable and whose island width is given by the steady state solution of (\ref{result}) with the substitution $\Delta'\rightarrow \Delta'+16 A_f/w^2$ and $\Sigma'\rightarrow \Sigma'+16A_f/w^2$.
Inspection of this equation, treating $\log{w}$ as a constant, shows that one can distinguish three different regimes according to the value of $\Delta'$. If $\Delta'\gg A_f^{1/3}$, the saturated island width is essentially the one given by the unforced case, i.e. $w_s\propto \Delta'$, with a modest increment due to the forcing. Near tearing mode marginal stability, $|\Delta'|\ll A_f^{1/3}$, $w_s\propto A_f^{1/3}$. The resulting island is thus much larger than would occur without an exterior perturbation. Finally, in the strongly stable case, $\Delta'\ll -A_f^{1/3}$, one finds $w_s\propto (-A_f/\Delta')^{1/2}$, previously given in Ref.~\cite{HK}. 
As regards the $\phi=\pi$ fixed points, the island width is given by the steady state solution of (\ref{result}) with the substitution $\Delta'\rightarrow \Delta'-16 A_f/w^2$ and $\Sigma'\rightarrow \Sigma'-16A_f/w^2$. The corresponding equation has physically acceptable ($w>0$) solutions, in the number of two, only if $\Delta'$ exceeds a positive critical value of  order $A_f^{1/3}$. These two fixed points are both unstable, the one with the largest width being a saddle point, unstable in the direction of the phase. Therefore, the island chain position always adjusts to be in phase with the external perturbation.

Equation \ref{result} has been cross-checked in two different ways. The analytical results have been obtained independently by two variants \cite{EPS,HMP} of  Thyagaraja's technique \cite{ Thya}. The numerical coefficients have been computed independently in \cite{HMP}.

As a conclusion, we have tackled the problem of nonlinear tearing
mode saturation by using a new rigorous perturbation expansion. Equation \ref{result} brings for the first time the correct expression up to terms of order $w^2$ for the island width evolution. A magnetic flux jump corresponding to a solenoidal effect in cylindrical geometry has been exhibited. The problem of the static forcing of a static plasma has been solved and displays three physical regimes. 
Preliminary calculations indicate that our perturbation technique works also in the case of a  velocity profile in the plasma, and of a forcing rotating at a possibly different speed than the resonant plasma region. Our approach can also easily be adapted to the case of cylindrical geometry and this will be presented elsewhere. It is important to appreciate that establishing solid analytic techniques to solve the classic
nonlinear tearing problem opens up new routes to deal with problems whose modeling goes
beyond conventional RRMHD. In particular, it would be interesting to revisit neoclassical and two-fluid diamagnetic effects on the generalized Rutherford equation. Last but not least, our
approach was not based on the specific $J=J(\psi)$ property but
rather on a straightforward perturbation expansion, and may
therefore be applicable to more complex physical models in which
$[\psi,J]\neq 0$, which will be done elsewhere. 

We acknowledge fruitful discussions with  J. Hastie, F. Militello and F. Porcelli which led to a joint oral contribution at the last IAEA meeting \cite{IAEA}.


\begin{thebibliography}{20}
\bibitem{Furth}
H.P. Furth, J. Killeen, M.N. Rosenbluth, Phys. Fluids {\bf 6}, 459 (1963)
.

\bibitem{Ruth}
P. Rutherford, Phys. Fluids {\bf 16},  1903  (1973).

\bibitem{White}
R.B. White {\it et al},  Phys.
Fluids {\bf 20}, 800 (1977);  L.E. Zakharov, {\it et al}, Sov. J. Plasma Physics {\bf 16}, 451 (1990); A. Pletzer and F.W. Perkins, Phys. Plasmas
{\bf 6}, 1589 (1999)
\bibitem{Thya}
A. Thyagaraja, Phys. Fluids {\bf 24}, 1716 (1981).

\bibitem{EO}
D.F. Escande, M. Ottaviani, Phys. Lett. {\bf A 323}, 278 (2004).

\bibitem{MP}
F. Militello  and F. Porcelli, Phys. Plasmas  {\bf 11}, L13 (2004)


\bibitem{Verhage}
A.J.L. Verhage et al., Nucl. Fusion {\bf 18} 457 (1978)

\bibitem{PPCF00}
D.F. Escande et al., Plasma Phys. Contr. Fus. {\bf 42}
B243 (2000)

\bibitem{HK} 
T. S. Hahm and R. M. Kulsrud, Phys. Fluids {\bf  28}, 2412 (1985) 


\bibitem{EPS}
N. Arcis, D.F. Escande, and M. Ottaviani, in {\it Proceedings of the 31$^{st}$ EPS Conference on Plasma Physics}, London UK, June 2004.

\bibitem{HMP}
R. J. Hastie, F. Militello  and F. Porcelli, submitted.

\bibitem{IAEA}
R. J. Hastie, F. Militello,  F. Porcelli, N. Arcis, D.F. Escande, and M. Ottaviani, paper PS/1-1, 20$^{th}$ IAEA Fusion Energy Conference, Vilamoura, November 2004.

\end{thebibliography}
\end{document}